\documentstyle[12pt,aaspp4]{article}  
\begin {document}
\title{{\bf Recent Dynamical Relaxation of Galaxy Clusters: Evidence for a Low-$\Omega_m$ Universe}}
\author{Adrian L. Melott\altaffilmark{1,3}, Scott W. Chambers\altaffilmark{1},
and Christopher J. Miller\altaffilmark{2}}
\altaffiltext{1}{Dept. of Physics \& Astronomy, Univ. of Kansas, Lawrence, KS 66045}
\altaffiltext{2}{Dept. of Physics \& Astronomy, Carnegie Mellon Univ., Pittsburgh, PA 15213}
\altaffiltext{3}{melott@kusmos.phsx.ukans.edu}

\newpage
\begin{abstract}
We argue that there has been substantial evolution in the
ellipticity of rich galaxy clusters between 0 $<$ z $<$ 0.1, and suggest that this
is additional evidence for a low matter-density Universe. 
\end{abstract}
\newpage

\section{Introduction}
Clusters of galaxies are often elongated systems (Carter \& Metcalfe 1980) whose
shapes and orientations are described by (projected) ellipticities and position angles.
Dynamical and cosmological information 
can be inferred from these rather general parameters.
For example, position angles show clusters are generally aligned 
with neighboring rich clusters, as suggested by projected galaxy distributions 
(Binggeli 1982; Rhee, van Haarlem \& Katgert 1992;
Plionis 1994) and X-ray isophotal morphologies 
({\it e.g.} West, Jones \& Forman 1995; Chambers, Melott \& Miller 2000, 2001).
Clusters tend to be aligned along these axes as a relic of gravitational
clustering.  However, if the flow slows greatly, as it will in a low-density
Universe, clusters will have time to relax to a more spherical shape
(e.g. Tsai \& Buote 1996; Buote 1998).
In a high-density cosmological background, mergers continue,
preventing the clusters form relaxing.

We investigate whether galaxy cluster ellipticity has evolved
significantly in recent cosmological times (0 $< z <$ 0.1).  
Clusters experiencing recent mergers will likely appear more elongated than relaxed clusters.

\section{Formulating The Null Hypothesis and Analysis}
In examining data related to the question of cluster alignment (Chambers,
Melott, and Miller 2001) we noticed that most of the signal for alignment
came from redshifts z $>$ 0.06.  We did not expect that this could be due
to incompleteness, since the surveys used there (MX Northern Abell Cluster Survey:
Slinglend {\it et al.} 1998; ESO Nearby Abell Cluster Survey:  Katgert {\it et al} 1996)
are 98\% complete in redshift measurements for Abell/ACO clusters of richness R $\ge$ 1
for redshifts 0 $<$ z $<$ 0.1.  Additionally, the number density as a function of
redshift is flat out to $z = 0.1$ (Miller \& Batuski 2001).
Furthermore, in that study we carefully compensated for any 
selection biases caused by the smaller
volume at low z by excluding from consideration any cluster whose nearest
neighbor was more distant than the nearest boundary of the survey region.
This assures that a ``true'' nearest neighbor is not missed,
lying just outside the sample.

One of the obstacles in alignment studies is random error in the
assignment of the position angle. 
If clusters had relaxed and become more
nearly spherical at cosmologically recent times, there would be correspondingly
greater ambiguity in the assignment of the axes.
Our null hypothesis is therefore that cluster ellipticity does not increase
with increasing redshift.  
The simplest approach with our two variables is to ask whether there is
a correlation between ellipticity and redshift.
 The linear correlation
coefficient tests for this and may be cast in the form
\begin{eqnarray} 
        r = \frac{\Sigma (\epsilon_{i}-\overline{\epsilon})(z_{i}-\overline{z}) }
                {\Sigma[(\epsilon_{i} -\overline{\epsilon} )^2(z_{i}-\overline{z})^2]^{1/2}}
\end{eqnarray} 
where the sum is over all $i$, and $\overline{\epsilon}$ and $\overline{z}$ are
the mean ellipticity and redshift, respectively.

\section{Sample Selection}
There are a number of issues that must be accounted for
when creating samples to be used in this analysis. First,
the base cluster catalog must show evidence for
being complete within the volume studied. This
catalog must also contain a large number of well-defined
clusters over a large portion of the sky. Projection effects
need to be minimized, and multi-wavelength data is a 
necessity.  Fortunately, the Abell/ACO catalog meets
all of these requirements. The current redshift data
for Abell/ACO clusters extends beyond $z=0.1$, covering
half of the sky (see e.g. Slinglend et al. 1998; Miller \&
Batuski 2001). If we restrict our catalog to
the richest Abell/ACO clusters, we can minimize
optical projection effects (Miller et al. 1999a,b).
A large number of Abell/ACO clusters have 
X--ray data, which can also be used to
avoid projection effects. There is substantial
evidence that these clusters are free of selection biases
and that they are a biased representation of the luminous mass
distribution  (Miller et al. 1999a; Miller \& Batuski 2001).

In order to reduce selection biases and assure accurate redshifts for our
sample, we excluded any clusters which not were present in the
range  0.012 $<$ z $<$ 0.1 and galactic latitude $|b| \ge$ 30$^o$.
We made a richness cut requiring $R \ge 1$ as specified in Abell (1958) or
Abell, Corwin, and Olowin (1989). This richness cut limits our studies
to the most massive and easily detected clusters.
However, $R=0$ clusters were never meant to be
treated as a complete sample of clusters and there are too many that are mis-identified
or un-identified, and so we exclude this subset entirely. 
We place a minimum physical size within
which ellipticities
are calculated at $1h^{-1}$Mpc. This is less than the
``typical'' Abell radius of $1.5h^{-1}$Mpc, but still
large enough to map the outer regions of clusters which
are most affected by infall.
The inner isophotes of
clusters are known to be less sensitive to the cosmological background
from simulations (Evrard 2001) and data (West \& Bothun 1990, hereafter WB).
We excluded any study having a circular ``mask'' within which counts or
measurements were made with radius $<$ 1 Mpc/h.  
Examples of such excluded datasets from the literature
include those of Carter \& Metcalfe (1980) and Binggeli (1982).
Binggeli notes that these methods systematically
underestimate ellipticity.

Rhee, van Haarlem \& Katgert (1991; RvHK hereafter) studied 107
rich R $\ge$ 1 cluster at z $<$ 0.1.  Using various methods, they
measured the ellipticity determined from galaxies 
within a radius $\sim$ 1-2 $h^{-1}$Mpc
of the cluster center.  We employ this sample, using the ellipticity 
measured by tensor method (RvHK).
The Kolokotronis et al. 2001 (KBPG) study of
22 rich cluster optical and X-ray images showed that their shape parameters
are generally correlated. Their optical
sample used a moments method on APM galaxies within r $\sim$ 1.8 $h^{-1}$Mpc
of the cluster center to determine the ellipticity.  
We also adopt this optical sample.
WB used a smoothing technique to
assign ellipticities to $\sim$ 70 clusters
in a substructure study.  We use the ellipticities determined from the 
outer regions in this study.

The use of X-ray samples is extremely important, due to a greatly reduced
possibility of projection effects contaminating ellipticity measurements.
Fewer X-ray samples are available. We use the McMillan, Kowalski \& Ulmer (1989; MKU)
sample of 49 {\it Einstein} X-ray clusters.
In their method, the outer regions of the cluster was emphasized.
We also study the KBPG sample of 22 ROSAT ellipticities, which focused on the central region
of the clusters.  Central regions tend to be rounder and therefore
contribute large random error to shape parameters. However,
KBPG demonstrated that the shape parameters from their optical 
and X-ray images are correlated.  Although the correlation is not
strong, we adopt their X-ray sample as a cross-check with the optical
sample. 
Again, we excluded any estimates of ellipticity based only on regions with
a radius $<$ 1 Mpc/h.
Plionis et al. (1991) described ellipticity evolution, but did not include
any X-ray data.

After being filtered through the richness, redshift, and galactic
latitude requirements described earlier, 
the RvHK optical sample is our largest sample with 96 clusters.
The MKU X-ray sample of 30 surviving clusters is the next largest.
WB reduces to 25 and KBPG to 18 clusters.
Later, WB will reduce further as ellipticities are not given for all
clusters.

\section{Analysis and Results}

We wish to determine whether ellipticity 
is positively correlated with redshift.
To this end, we test our samples for linear correlation
(Bevington and Robinson 1992).
The correlation coefficient, $r$ has a range of values  $-1 < r < 1$.
If $|r|$ is close to 1, the data are strongly correlated.  If $r$=0 the data are
uncorrelated.  We therefore expect $r >$ 0 if clusters are more elliptical at higher
redshifts, $r$=0 if there is no change, and $r <$ 0 if ellipticity decreases
with redshift.
We have not merged the samples because of apparent systematic differences
in ellipticity assignment.  Authors used different methods
to measure ellipticity.   
WB systematically excluded ellipticities which they did not conclude were significant,
and we see systematically larger ellipticities with a strong redshift
gradient.
Merging samples with very different mean ellipticities
may falsely kill or intensify the signal. 
We have, however, converted all ellipticities given to the same notation,
\begin{eqnarray}
        \epsilon=1-\frac{b}{a}
\end{eqnarray}
where $a$ is the projected major axis and $b$ is the projected minor axis.
We note that we are dealing with projections and that intrinsically more
spherical systems will more often appear circular in projection (Plionis et al. 1991).

The results of the correlation statistic are given in Table 1.
Included is the sample name, number of clusters, $N_{cl}$, 
and the correlation coefficient, $r$.
The last column contains the probability $P$ that the correlation
coefficient could arise from an uncorrelated sample.
The confidence is 1 - $P$.
Interestingly, for all samples, $r$ is positive, ranging from 0.22 to 0.67, 
indicative of cluster ellipticity being an increasing function of redshift.
The range of values reflects systematic differences in the samples.

MKU are the only authors to provide errors on their ellipticities.
For consistency, we have given in the table an unweighted $r$.
However, the error-weighted $r$ for this sample is 0.39, essentially the
same as the unweighted.
The WB sample has the largest correlation coefficient with highest confidence.
The ellipticities in the WB sample are
higher and increase more rapidly with redshift than the other samples.
This, and the small scatter in the data, results in WB's strong signal.
We note that WB applied considerable smoothing to cluster profiles, and
rejected any results with high uncertainty, when computing ellipticities.
In Figures 1 \& 2 we plot the ellipticity vs. redshift for the X-ray and optical
samples, respectively.
The increase with redshift is apparent.
It is noteworthy
that the increasing trend exists in all X-ray and optical samples.

\section{Possible Sources of Systematic Error}
We cannot exclude some type of systematic error as the source of this
signal.  However, we have removed some of the more obvious possibilities
by keeping only those clusters which are members of a carefully controlled
volume-limited sample and by using both optical and X-ray
samples.
Our results depend on accurate measurements of the
cluster ellipticities. Confirmation of our
results will be possible using upcoming controlled cluster
catalogs from the Sloan Digital Sky Survey  (SDSS)
(e.g. Kim et al. 1999; Annis et al. 1999).  The five color 
photometry of the SDSS data will enable accurate ellipticity
calculations for clusters to $z \sim 0.4$ (see Kim et al. 2000).

{\footnotesize
\begin{deluxetable}{ccccc}
\tablewidth{0pt}
\tablecaption{Ellipticity-Redshift Correlation}
\tablehead{
\colhead{}  & \colhead{}  & \colhead{}  & \colhead{}  & \colhead{} \\
\cline{1-5} \\ 
\colhead{Sample} &  \colhead{} &\colhead{N$_{cl}$} & \colhead{r} & \colhead{Prob.} }
\startdata
KBPG      &  opt.       &18     & 0.59  &2 e -3\nl
KBPG      &  xray       &18     & 0.31  &1 e -1\nl
MKU       &  xray       &30     & 0.41  &9 e -3\nl
RvHK      &  opt.       &96     & 0.22  &1.4 e -2\nl
WB        &  opt.       &18     & 0.67  & 1.5 e -4\nl
\enddata
\end{deluxetable}}

In the meantime, we can address some more obvious effects which
may lead to a false signal.  For instance, 
clusters of a fixed physical size at higher redshift are smaller in angular size. It is
therefore more difficult to measure ellipticities as they appear on
the sky. However, at our redshift limit of $z=0.1$, a typical cluster diameter
is still $\sim 20-30$ arc-minutes.
Therefore, ellipticity measurements should
not be unduly difficult within our specified redshift range. 

Another possible source of contamination that could affect our samples is
redshift bias in the distribution of cooling flow (CF) clusters.
Cooling flow clusters tend to be more relaxed and spherical in the X-ray
but need not be so in the optical (Henriksen 1993).  Thus we need to address
the possibility that our X-ray samples could be biased if the distribution
of CF clusters were biased by redshift.
Using the compilation from Loken, Melott, and Miller (1999),
we find that the number of identified cooling flow clusters within our
X--ray samples is 40\% and 17\% for the MKU and KBPG samples
respectively. We performed a Kolmogorov-Smirnov test on the distribution
of redshifts for the MKU cooling flow clusters versus the total sample and
find no significant difference.  In fact, the K-S test indicates a 95\% 
probability that the distribution of CF clusters in the MKU sample could have
been drawn from the parent population of that entire sample.
The KBPG contains too few cooling flow clusters
to perform a K-S test. However, the three KBPG cooling flow clusters
have redshifts z=0.046, 0.0602, and 0.0750, and so they cannot be contributing
to our signal.

We excluded clusters 
for which WB conclude the ellipticity is not
measurable due to large error. A large fraction of 
these unmeasurable clusters at our redshift limits could 
also induce a false signal. 
WB do not specify errors or their exact procedure for rejecting ellipticities.
The 7 clusters excluded out of 25 which survived our cuts are at small to
moderate redshift.
We assess the significance of their culling by re-introducing the
rejected clusters with mock ellipticities.  If we assign them ellipticities
equal to the mean in their redshift bin (binsize 0.01) for all the other
samples, $r$ increases to 0.79 with a probability of no correlation $< 1 e -8$. 
If we assign them ellipticities equal to the mean of the WB for all redshifts,
$r$ decreases to 0.40 with a probability of $ 4 e -2$.
The latter procedure would mimic the effect if they had systematically
excluded clusters with high ellipticity at moderate redshift.

There are, of course, other biases that could exist, such as in the
luminosity selection of the subsamples. 
We note that de Theije et al. (1995) in a re-analysis of ellipticity studies,
argue that most methods underestimate large ellipticities, and overestimate
small ones, so that the evolution could be even faster than we have measured.
They did not include WB in this study.
Differences in methods could account for the differences in $r$ values between
various samples used here.

\section{Discussion}
The hypothesis of evolving ellipticity was formulated on a simple physical
basis.  Perturbations are typically highly anisotropic when they first 
collapse (Shandarin et al. 1995).  Gravitational clustering in the nonlinear
regime results in the flow of matter along filaments and sheets to form
clusters (Shandarin and Klypin 1994).  These clusters would then become
part of a larger-scale flow and merging to produce even larger objects
(Pauls and Melott 1995).  

Once the clusters are essentially collapsed along all 3 axes, relaxation
can begin.  Gravitational relaxation is well-understood to move toward
isotropizing the system.  Thus, if
the clusters were isolated, they would become more spherical (circular in
projection).

In a high-density Universe, mergers
are expected to continue to the present time.
In a low-$\Omega_m$
Universe, with or without a cosmological constant, linear gravitational
clustering shuts down as the matter density declines (Peebles 1980;
Lahav et al. 1991; Hamilton 2001).

If clustering continues unabated, new mergers and infall of matter along
filaments will tend to keep clusters aspherical.   If linear clustering
shuts down, mergers will tend to run out of fuel after existing bound
objects have completed their infall.  Our signal for ellipticity evolution
is additional evidence for the kind of isolation and relaxation of clusters
that will be characteristic of a low-density Universe.

Is the timescale reasonable?
Relaxation can take place as a consequence of the time-dependent gravitational
potential  which varies on a crossing time.  It
is therefore appropriate to compare the crossing time for clusters to the
time over which we observe evolution.

The crossing time for a galaxy in a cluster is
\begin{eqnarray}
        t_{cross} \sim (\frac{R}{1Mpc})(\frac{1000km/s}{\sigma_v})Gyr.
\end{eqnarray}
where $R$ is the radius in Mpc and $\sigma_v$ is the line of sight velocity dispersion
in km/s.
This can be compared with the roughly 1.3 Gyr timescale for our sample, assuming
h=0.7, $q_0$=0.
This interval is also comparable to the time gas remains in dis-equilibrium
after having been stripped by the merger of two groups (Evrard 1990).

Further supporting evidence
comes from an analysis of the Butcher-Oemler effect (Ellingson et al.
2001) arguing that the effect is dominated by a redshift-dependent
number of blue galaxies in clusters, which are primarily located in the
outer regions of these clusters, and show dramatic declines in infall
rates.  Wang \& Ulmer (1997) note a strong correlation between cluster
ellipticity and blue fraction, further supporting this connection.

Our data have too much
scatter to justify going beyond linear order.  Future work with larger,
more uniform samples may enable a detailed comparison with
theory and ellipticity evolution in simulations.
In particular, we look forward to testing this result against (1) X-ray selected cluster
samples sensitive to extended emission (2) Optically selected catalogs based on new selection
techniques which will produce even more uniform samples, based on surveys
such as SDSS and 2dF (York et al. 2000; Davis \& Newman 2001).

\section{Acknowledgments}
We are grateful for support from the NSF under grant number
AST-0070702, as well as useful comments from the referee
Andrew Hamilton.

\begin{figure}[t]
\epsscale{.7}
\plotone{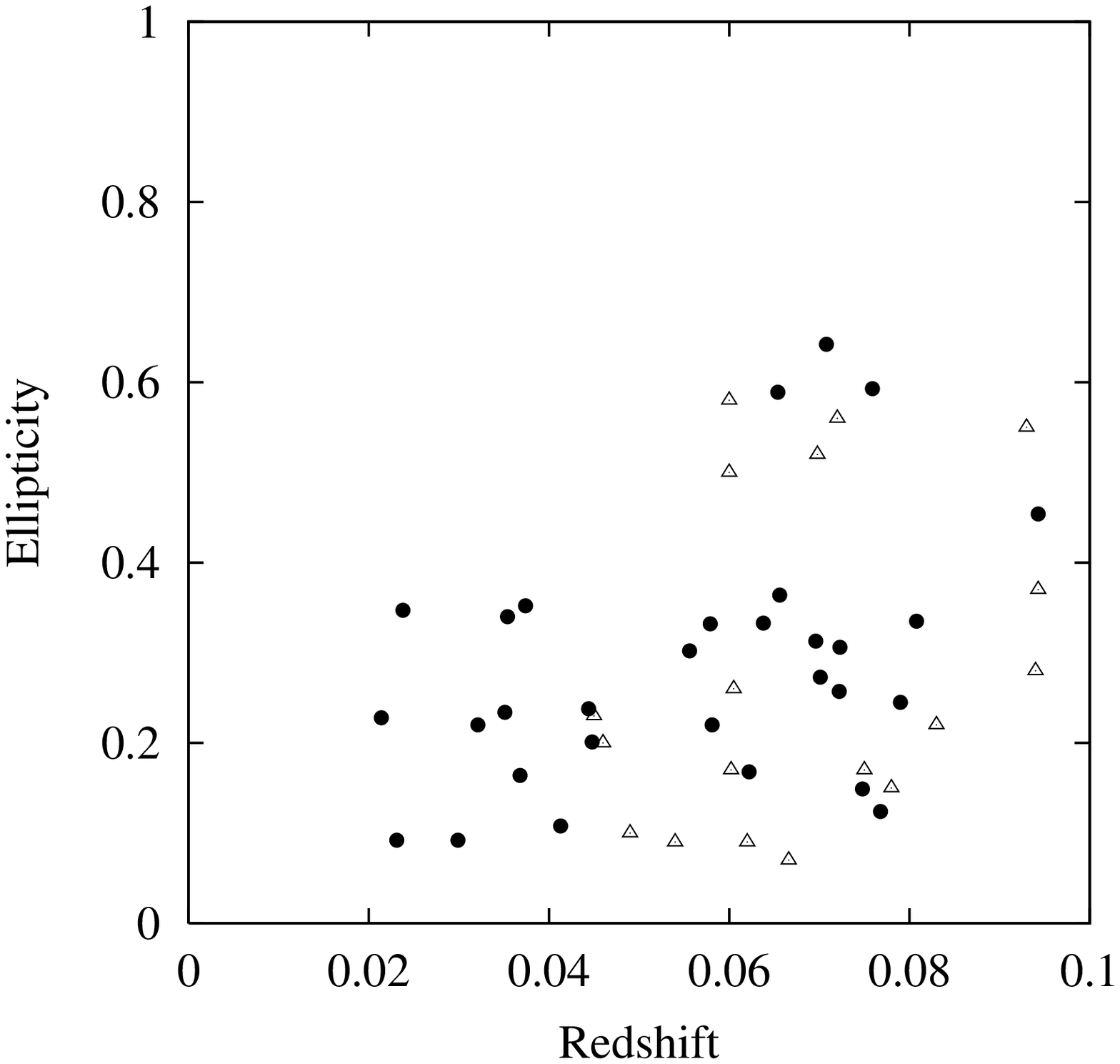}
\caption[]{\footnotesize
Ellipticity vs. redshift for X-ray samples.  Filled circles
represent MKU data; empty triangles are KBPG.
}
\end{figure}

\begin{figure}[t]
\epsscale{.7}
\plotone{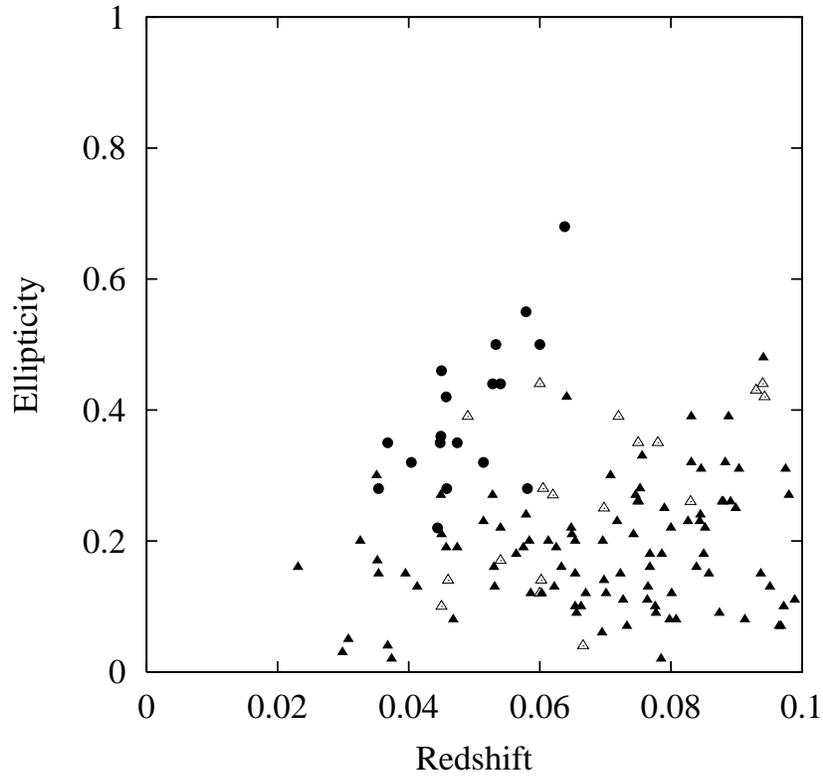}
\caption[]{\footnotesize
Ellipticity vs. redshift for optical samples.  Filled circles
are WB data; filled triangles are RvHK data; 
empty triangles are KBPG.
}
\end{figure}

\end {document}